\documentclass[11pt]{article}\topmargin=-0.5cm\hfuzz=10pt  \oddsidemargin=-0.1cm \textheight 222mm\textwidth=16.3cm
\usepackage{amssymb}
\newcommand{\comm}[1]{}

\def\short{\comm}

\usepackage{color}

\usepackage{endnotes}

\newtheorem{theorem}{Theorem}
\newtheorem{lemma}{Lemma}
\newtheorem{proposition}{Proposition}
\newtheorem{definition}{Definition}
\newtheorem{remark}{Remark}
\newtheorem{example}{Example}

\def\e{\varepsilon}

\def\defi{\stackrel{{\scriptscriptstyle \Delta}}{=}}
\def\defi{:=}

\def\a{\alpha}
\def\d{\delta}
\def\o{\omega}
\def\O{\Omega}

\def\F{{\cal F}}
\def\w{\widehat}
\def \Ind{{\,\rm Ind\,}}
\def \Ind{{\mathbb{I}}}

\def\mes{{\rm mes\,}}

\def\esssup{\mathop{\rm ess\, sup}}

\def\Re{{\rm Re\,}}
\def\Im{{\rm Im\,}}

\def\R{{\bf R}}

\def\Z{{\cal Z}}

\def\C{{\bf C}}

\def\ww{\widetilde}
\def\X{{\cal X}}

\def\oo{\bar}

\def\GG{{\cal G}}
\def\U{{\cal U}}
\def\V{{\cal V}}

\def\M{{\cal M}}

\def\CC{{\cal A}}

\def\WO{\stackrel{\circ}{{W}}}

\newcommand{\be}{\begin{equation}}
\newcommand{\ee}{\end{equation}}
\newcommand{\bd}{\begin{displaymath}}
\newcommand{\ed}{\end{displaymath}}
\newcommand{\ba}{\begin{array}{ll}}
\newcommand{\ea}{\end{array}}
\newcommand{\baa}{\begin{eqnarray}}
\newcommand{\eaa}{\end{eqnarray}}
\newcommand{\baaa}{\begin{eqnarray*}}
\newcommand{\eaaa}{\end{eqnarray*}}

\def\oo{\bar}

\def\a{\alpha}

\def\ew{\left(e^{i\o}\right)}
\def\eww{\left(e^{i\cdot}\right)}

\def\ZZ{{\mathbb{Z}}}

\def\TT{{\mathbb{T}}}

\def\g{\gamma}
\def\G{\Gamma}
\def\ee{\e}
\def\WO{\stackrel{p}{W_2^1}}

\date{Submitted:  October 16, 2023. Revised: June 4, 2024}
\title{
Spectral representation of two-sided signals from  $\ell_\infty$ and applications to  signal processing 
}
\author{Nikolai Dokuchaev}
 
\begin{document}
 \def\short{\comm}
\def\brea{}
\def\breakk{}
\def\break{}
\def\BRR{}
\def\breakm{\nonumber\\  }\def\BR{}\def\BRR{}
\def\breacm{}
\def\dt{}
\def\ZZM{\ZZ_{\scriptscriptstyle{\le 0}}}
\maketitle
 \begin{abstract} The paper studies   spectral representation as well as 
 predictability and recoverability problems for  non-vanishing  discrete time signals
 from $\ell_\infty$, i.e. for bounded discrete time signals, including signals that do not vanish at  $\pm\infty$. The extends the  notions of transfer functions, the spectrum gaps, bandlimitness,  and filters,
  on these  general type signals. Some frequency conditions of predictability  and data recoverability are presented, and some recovery methods and predictors have been suggested.  \end{abstract} 

\comm{discrete time signals, spectral representation,  transfer functions, filters,   data recovery, predicting} 
\section{Introduction}
The most important  tools  used for signal processing and system theory are based on the representation of signal processes
in the frequency domain. This includes, in particular,   the notions of transfer functions, spectrum gaps, 
filters, conditions of predictability  and  data recoverability. 
 For the continuous time processes $x(t)|_{t\in\R}$, the spectrum  representation is via  the Fourier transform for two-sided processes vanishing as $t\to\pm\infty$ 
and via the  Laplace transform for one-sided processes being zero on a half of the time axis
but not necessarily vanishing on the other half of the time axis.  A similar situation is for the discrete time processes $\{x(t)\}_{t=-\infty}^{+\infty}$ and their spectrum representation  via  Z-transform.
For two-sided processes vanishing sufficiently fast 
on $\pm\infty$ such as  processes from $\ell_2$, this Z-transform is 
well defined on the unit circle $\{z\in\C:\ |z|=1\}$. 
For one-sided processes  $\{x(t)\}_{t=-\infty}^{+\infty}$ from $\ell_\infty$, i.e. such that either  $x(t)=0$ for $t<0$
or  $x(t)=0$ for $t>0$, one can apply Z-transform defined in some open domains with circular boundaries  either outside
or inside of the unit circle $\{z\in\C:\ |z|=1\}$.
 In this case, the signals they  do not have to vanish on the other half of the time axis. 
For the special case of  bounded  one-sided processes  $\{x(t)\}_{t=-\infty}^{+\infty}$, i.e. such that either  $x(t)=0$ for $t<0$ or $x(t)=0$ for $t>0$, this Z-transform is defined  either in the domain 
$\{z\in\C:\ |z|<1\}$ or in the  domain 
 $\{z\in\C:\ |z|>1\}$. 

 It can be observed that any signal from $\ell_\infty$ can be modified  to a signal from $\ell_1$ without any loss of information, for example, by replacement $x(t)$ by $e^{-|t|}x(t)$.  However,  at least for the case of  signals from $\ell_2$, these damping transformations represent  the convolutions on 
the circle $\{z\in\C:\ |z|=1\}$ in the frequency domain,  with smoothing kernels. Unfortunately, 
these transformations  would remove spectrum degeneracies  commonly exploited in data recovery and prediction for signal processing.  For the general type  two-sided processes  from $\ell_\infty$, one could expect a similar impact of the damping transformations on the spectrum.
 This
could be inconvenient, since it 
imposes undesirable restrictions on the underlying models.
This is our motivation for studying  spectral representation for general type non-vanoshing bounded signals.

Formally,  the spectral representation  for bounded non-vanishing discrete time signals  is defined 
as  Fourier transforms for pseudo-measures on $[-\pi,\pi]$ were represented as elements of $\ell_\infty$; see  Chapter III in \cite{Kahane}.  However, this definition didn't lead so far to 
frequency based notions and methods such as filtering, predicting, and data recovery, for non-vanishing signals. 
 The paper  suggests more constructive definition of  spectral representation for  non-vanishing signals. Based on this definition, the paper extends the  notions of transfer functions, spectrum gaps,  and filters, on these
general type signals (Section  \ref{SecA}).  This allowed to obtain some frequency conditions of predictability  and  data recoverability for non-vanishing signals with spectrum degeneracy (see Section \ref{SecRkn}).
\par
It can be noted the a similar approach  was developed for spectral representation and predicting of 
non-vanishing bounded continuous time signals  in \cite{D24}. This spectral representation  was applied for the sampling problem   and interpolation formula in \cite{D24s}.

\subsection*{Some notations}

Let $\ZZ$, $\R$, and $\C$, be the set of all integer, real, and complex numbers, respectively.

\def\DD{\mathbb{D}}

Let $\TT\defi\{z\in\C:\ |z|=1\}$,  $\DD\defi  \{z\in \C: \ |z|>1\}$, and $\oo\DD\defi  \{z\in \C: \ |z|\ge 1\}$.

We denote by $\ell_\infty$ the set of all processes (signals) $x:\ZZ\to \C$, such that
$\|x\|_{\ell_\infty}\defi \sup_{t\in\ZZ}|x(t)|<+\infty$.

For $r\in[1,\infty)$, we denote by $\ell_r$ the set of all processes (signals) $x:\ZZ\to \C$, such that
$\|x\|_{\ell_r}\defi \left(\sum_{t=-\infty}^{\infty}|x(t)|^r\right)^{1/r}<+\infty$.

Let $C([-\pi,\pi])$ be   the standard linear  space of continuous functions $f: [-\pi,\pi]\to\C$
with the uniform norm $\|f\|_C\defi \sup_\o |f(\o)|$.

Let  ${\WO}(-\pi,\pi)$ denote  the Sobolev  space of functions $f: [-\pi,\pi]\to\C$
that belong to $L_2(-\pi,\pi)$ together with the distributional derivatives
up to the first order, and such that $f(-\pi)=f(\pi)$.

\par

We denote by $\Ind$  the indicator function.

\section{Spectral representation of processes from $\ell_\infty$}
\label{SecM}

Let  $\CC$  be  the space of functions $f\in C([-\pi,\pi])$
with the  finite norm $\|f\|_{\CC}\defi \sum_{k\in\ZZ}|\w f_k|$, where 
$\w f_k=\frac{1}{2\pi}\int_{-\pi}^\pi e^{-i\o s}f(s)ds$ are the Fourier coefficients of $f$.
In other words, $\CC$ is the space of absolutely convergent Fourier series on $[-\pi,\pi]$. 
By the choice of its norm, this is a separable Banach space that is isomorphic to $\ell_1$. 

Clearly, the embeddings 
 $\CC\subset  C([-\pi,\pi])$ is continuous.
It can be noted that there are functions  in $C([-\pi,\pi])$ that do not belong to $\CC$; see, e.g.,
\cite{Katz}, p.113.

We assume that each $X\in L_1([-\pi,\pi])$ represents an element of the dual space  $C([-\pi,\pi])^*$ such that
 $\langle X,f\rangle=\frac{1}{2\pi}\int_{-\pi}^\pi X(\o)f(\o)d\o$ for $f\in C([-\pi,\pi])$. We will use the same notation
  $\langle \cdot,\cdot\rangle$ for the extension of this bilinear form  
  on  $\CC^*\times \CC$.

 \begin{proposition}\label{prop1} 
\begin{enumerate}
\item
If $f\in \CC$ and $g\in\CC$, then $h=fg\in\CC$, and $\|h\|_{\CC}\le \|f\|_{\CC}\|g\|_{\CC}$.   
\item The embedding 
$\WO(-\pi,\pi)\subset\CC$ is continuous.
\end{enumerate}
\end{proposition}

It follows that  the embeddings 
\baaa
\CC \subset C([-\pi,\pi]) \subset L_1([-\pi,\pi]),
\qquad L_1([-\pi,\pi])^*\subset  C([-\pi,\pi])^* \subset \CC^*
\eaaa 
are continuous.  

The space $\CC$ and its dual $\CC^*$ can be used to define formally a  spectral representation 
for $x\in \ell_\infty$ via $X\in\CC^*$  
such that 
$\langle X,f\rangle=\sum_{t\in\ZZ}x(k)\w f_k $ for any $f\in \CC$, where $\{\w f_k\}\in\ell_1$  is the series of the Fourier coefficients for $f$, similarly to  Chapter III in \cite{Kahane}, where the Fourier transforms for pseudo-measures on $[-\pi,\pi]$ were represented as elements of $\ell_\infty$.   For the continuous time signals,   a definition of the Fourier transform  via a similar  duality 
is given in Chapter VI in \cite{Katz}. 
However, for the  purposes related to the problems of recoverability and prediction of digital signals, we will need a straightforward  definition based on the following lemma. 

\begin{lemma}\label{lemma1}  For any $x\in\ell_\infty$, there exists a 
weak* limit   $X\in \CC^*$ of the sequence of functions 
$X_m(\o)\defi \sum_{t=-m}^m e^{-i\o t} x(t)$ defined on $[-\pi,\pi]$ for $m=1,2,...$.
This $X$ is such that $\|X\|_{\CC^*}=\|x\|_\infty$ and that 
$\langle X,f\rangle=\sum_{t\in\ZZ}x(k)\w f_k $, where $\{\w f_k\}\in\ell_1$  is the series of the Fourier coefficients for $f$.  
\end{lemma}

 It can be noted that, in the lemma above,  $X_m\in L_1([-\pi,\pi]) \subset C([-\pi,\pi])^*\subset \CC^*$.

 We define a  mapping  $\F:\ell_\infty\to \CC^*$  such that  $X=\F x$ for $x\in\ell_\infty$ is  the  limit  in $\CC^*$  introduced in Lemma \ref{lemma1}. By  Lemma \ref{lemma1}, this mapping is linear and continuous.

 Further, define a mapping  $\GG:\CC^*\to \ell_\infty$ such that
\baaa
x(t)=\langle X,e^{i\cdot t}\rangle \quad \hbox{for}\quad  x=\GG X,\quad X\in \CC^*, \quad t\in\ZZ.
\eaaa
Clearly, the operator $\GG: \CC^*\to\ell_\infty$ is linear and continuous. In addition, for $x=\GG X$ and  $x_m\defi \GG X_m$, we have that $x_m(t)=x(t)\Ind_{|t|\le m}$.

\begin{theorem}\label{Th1} The mappings  $\F:\ell_\infty\to\CC^*$ and 
$\GG:\CC^*\to \ell_\infty$ are continuous isometric bijections such that 
$\F=\GG^{-1}$ and  $\GG=\F^{-1}$. 
\end{theorem}
Based on this, we will write  $\F^{-1}$ instead of $\GG$.

 \begin{remark} The space $\CC$ was selected by the following reasons:
 \begin{enumerate}
 \item
 it is wide enough, with weak enough topology,  to embed the set of functions $\{e^{i\cdot t}\}_{t\in\ZZ}$ in its unit ball, and
 \item it is tiny enough, with  strong  enough topology, to ensure that Lemma \ref{lemma1} holds.
 \end{enumerate} 
\end{remark}

Further, we say that $X\in\CC^*$ is real valued (imaginary valued) if $ \langle X,f\rangle$ is real (or imaginary) for any real valued $f\in\CC$. 
 Clearly, any $X\in\CC^*$ allows a  unique representation $X=\oo X+i \ww X$, where $\oo X,\ww X\in \CC^*$ are  real. We will use notations $\Re X$ for $\oo X$ and $\Im X$ for $\ww X$.     
 
 We say that a real valued $X\in\CC^*$ is odd (even) if $ \langle X,f\rangle=0$ for any real valued  odd (even) $f\in\CC$.
 
 It is easy to show that if $X\in\CC^*$ is real valued then $\Re X$ is even and $\Im X$ is odd;
if $X\in\CC^*$ is real valued and even then  $\Im X=0$.
\section{Applications for signal processing}\label{SecA}
The spectral representation introduced above allows to extend some standard tools of signal processing on the case of signals from $\ell_\infty$. In particular,  it allows to characterise   signals from $\ell_\infty$ featuring spectral gaps, such as band limited processes. 
Also, it  supports implementation of important  tools such as transfer functions and low-pass filter, and it helps to obtain predicting and data recovery algorithms for these signals.

\subsection{Transfer functions}
The existing theory does not consider  transfer functions applied to the 
general type two-sided processes $x(t)$ from $\ell_\infty$ that do not vanish  $t\to \pm\infty$.  The suggested above spectral representation of  $x\in\ell_\infty$  via elements from $\CC^*$ allows to implement the transfer functions for   all $x\in\ell_\infty$.  
 
\begin{definition}\label{defTF}
Let $H:\TT\to\C$ be  such that the function $H\left( e^{i\cdot}\right)$ defined 
on $[-\pi,\pi]$ belongs to $\CC$. Then we say that $H\left( e^{i\cdot}\right)$
is a transfer function on $\ell_\infty$. For $x\in\ell_\infty$ and $X=\F x$, we define 
$\w  X=H X\in \CC^*$ and $\w x=\F^{-1} \w X$ 
 such that $\langle \w X,f\rangle\defi \langle X,H\left( e^{i\cdot}\right) f\rangle $. 
  \end{definition}
  It can be noted that,   by Proposition \ref{prop1}(i), for any $f\in\CC$, we have that 
   $H\left( e^{i\cdot}\right)f\in\CC$ as well. Hence $\w X=HX$ is well defined as an element of $\CC^*$.
    \begin{lemma}\label{lemmaTF}
Let $H:\TT\to\C$ be  such that $H\left( e^{i\cdot }\right)\in\CC$, and let  $H\left( e^{i\o}\right)=\sum_{k\in\ZZ}\w h_ke^{i\o k}$,
where $\{\w h_k\}_{k\in\ZZ}\in\ell_1$. Let $X\in\CC^*$, $\w X=HX$, and $\w x=\F^{-1} \w X$. Then
\baaa
\w x(t)=\sum_{q\in\ZZ}\w h_{t-q}x(q).
\eaaa
The series absolutely converges uniformly over any bounded set of  $x\in \ell_\infty$.
  \end{lemma}

  \subsection{Spectrum degeneracy}
The spectral representation introduced above allows to describe   signals from $\ell_\infty$ featuring spectral gaps, such as band limited processes, as well as weaker types of spectrum degeneracy.

 \begin{definition}\label{defPointDeg}
 \begin{itemize}
\item[(i)]
For a Borel measurable set  $D\subset [-\pi,\pi]$ with non-empty interior,   let  
$x\in\ell_\infty$ be such that  
 $\langle  \F x,f\rangle =0$ for any $f\in \CC$ such  that $f|_{[-\pi,\pi]\setminus D}\equiv 0$.
 In this case, we say that $D$ is a spectral gap of $x\in\ell_\infty$ and of $X=\F X$.
 \item[(ii)]
Let $\G$ be a set.  Assume that a function $G:\R\times \G\to\C$ 
be such that $G(\cdot,g)\in\CC$ for each $g\in \G$, and $\sup_{g\in \G}\|G(\cdot,g)\|_\CC= +\infty$.
Let $x\in \ell_\infty$ be such that $\sup_{g\in \G}\|G(\cdot,g)X\|_{\CC^*}<+\infty$.  
Then we say that the signal $x$ features spectrum degeneracy  that compensates   $G$.
\end{itemize}
\end{definition}

\begin{example}\label{ex1} Let $r>0$, $\G=(0,1)$, and  
$G(\o,\nu)= (|\o|^r+\nu)^{-1}$. Let $x\in \ell_1\cap \ell_\infty$ be such that   $\esssup_{\o\in[-\pi,\pi]}|X(\o)|/G(\o,0)<+\infty$ for $X=\F x$. Then $x$
 features spectrum degeneracy  that compensates   $G$. 
 \end{example}
In the example above, $G(\cdot,\nu)\in \WO([-\pi,\pi)$, hence  $G(\cdot,\nu)\in \CC$ for $\nu>0$. The corresponding spectrum degeneracy is a single point spectrum degeneracy.

 \subsection{Filters}
Unfortunately, the ideal low-pass and high-pass filters with rectangle profile of the transfer function do not belong to $\CC$. Hence they are not covered by 
Definition \ref{defTF} of the transfer functions  applicable to signals from $\ell_\infty$.   However,  some approximations of these ideal filters can be achieved  with trapezoid response functions from $\CC$. For  example, let 
\baaa
H_{p,q}\left( e^{i\o}\right)\defi \Ind_{\{\o\in [-p,p]\}}+
\frac{q-\o}{q-p}\Ind_{\{\o\in (p,q]\}} 
+\frac{q+\o}{q-p}\Ind_{\{\o\in [-q,-p)\}},
\eaaa
 where $0<p<q<\pi$. Since $\WO(-\pi,\pi)\subset \CC$, the functions  $H_{p,q}\left( e^{i\cdot}\right)$ belong to $\CC$, hence  they are admissible  transfer functions on $\ell_\infty$.
 Clearly, for any $x\in \ell_\infty$ and $X=\F x$, we have that $H_{p,q}X$ has a spectral gap on 
 $(-\pi,-q)\cup (q,\pi)$; in this sense, the filtered  process $\w x=\F^{-1} (H_{p,q}X)$ is  band-limited.
 
  For $q\to p+$, these functions  approximate the ideal  low pass filter with the pass interval $(-p,p)$, i.e. with the rectangular transfer function  $H_{p,p}\left( e^{i\o}\right)=\Ind_{\{\o\in [-p,p]\}}$. As was mentioned above,  $H_{p,p}\left( e^{i\cdot}\right)\notin \CC$, hence it does not represent a transfer function   that is applicable for signals from $\ell_\infty$.

\begin{remark}\label{rem1} For any $p>0$ and $q>p$ and for  $\G=\{\nu:\ \nu>0\}$,  each signal $x$ with a spectrum gap $[-q,q]$ features spectrum degeneracy  that compensates   $G(\o,p,q)=\nu H_{p,q}(\o)$.
\end{remark}

\subsection{Causal transfer functions}
\begin{definition}\label{defCas}
We  say that a transfer function $H$ on $\ell_\infty$ such as described in Definition \ref{defTF} is causal if, for any $\tau\in\ZZ$ and any $x\in \ell_\infty$ such that if
$x(t)=0$ for all $t\le\tau$ then $\w x(\tau)=0$,  
  where $\w x=\F^{-1} \w X$ and $\w  X=H X$. 
 \end{definition}
It can be noted that, in the definition above, $\w  X\in \CC^*$. 

\begin{theorem}\label{ThC}
Assume that a function $H:\oo\DD\to\C$ is continuous on $\oo\DD$ and analytic on $\DD$, and that 
$H\left( e^{i\cdot}\right)\in \CC$. Then $H\left( e^{i\cdot}\right)$ 
is a transfer function on $\ell_\infty$ that is causal in the sense of Definition  \ref{defCas}.
  \end{theorem}

 \section{Applications for data recovery and prediction problems} 
 \label{SecRkn}

 
 We consider the task of recovering of non-observed values $x(t_k)|_{k\in\M}$ from the observed values $x(t_k)|_{k\in\ZZ\setminus\M}$  for signals from certain subsets of $\ell_\infty$. 

\subsection{Recovery of a finite set of missed values} 
 
 Let $D\subset [-\pi,\pi]$ be a Borel set with non-empty interior.   Let $\V(D)$ (or $\V_R(D)$, or  $\V_I(D)$) be the set of all signals $x\in\ell_\infty$ such that 
 $x= x_v+x_s$, where 
 \begin{itemize}
 \item
 $D$ is a spectral gap for $X_v=\F x_v$ (or for   $\Re X_v$, or for $\Im X_v$, respectively);
 \item $X_s=\F x_s$ is such that $X_s\in C([-\pi,\pi])^*$, and  this $X_s$ is represented  by a Radon measure on $[-\pi,\pi]$ that is singular with respect to the Lebesgue measure.
  \end{itemize} 
In particular,  the corresponding signals include $x_s$ include signals   $\sum_{k\in\ZZ}\a_ke^{i \o_k t}$ for all  $\{\a_k\}_{k\in\ZZ}^\infty \in\ell_1$. In this case,  $X_s(\cdot )=\sum_{k\in\ZZ}\a_k\d(\cdot-\o_k)$,
where $\d(\cdot-\o_k)$ are delta-functions, i.e.
$\langle  \d(\cdot-\o_k), f\rangle =f(\o_k)$ for $f\in C([-\pi,\pi])$. 

Clearly, $\V(D)\subset \V_R(D)\cap \V_I(D)$.

 \begin{theorem}\label{ThR} For any Borel set $D\subset [-\pi,\pi]$ set with non-empty interior, for any finite set  $\M\subset\ZZ$,  the values 
$x(t)|_{t\in\M}$ for any $x\in\V_R(D)\cup \V_I(D)$
are uniquely defined by the values $x(t)|_{t\in\ZZ\setminus\M}$.
\end{theorem}

As far as we know,  the impact on the recoverability of the degeneracy featured only by the real (imaginary) part of the signal spectrum  
 has not been presented in the existing literature.

Formally, Theorem \ref{ThR} implies a method of recovery for $x|_\M$ since the trigonometric polynomial  
$X_\M(\o)=\sum_{t\in\M}e^{-it\o}x(t)
$  is observable on $D$ in the following sense: for any $f\in\CC$  supported on $D$, we have that $\langle X_\M,f\rangle=- \langle X_{\ZZ\setminus\M},f\rangle$, where  
$X_{\ZZ\setminus\M}\defi \F (\Ind_{\cdot \notin\M}x(\cdot))$.  However, this would require to calculate  $X_{\ZZ\setminus\M}$, which seems to be numerically challenging.
The following theorem gives  an alternative approach based on implementation  of explicitly given causal transfer functions and applicable for  signals from a more narrow class $\V(D)$.  
 \subsection{Predicting problem} 
Let $c>0$ and $q>1$ be given. For $\w\o\in (-\pi,\pi]$,  $\nu\in (0,1)$, let \baaa
G(\o,\w\o,\nu)\defi \exp\frac{c}{|e^{i\o}-e^{i\w\o}|^q+\nu}.
\eaaa 
The functions   $G(\cdot,\w\o,\nu)\in \WO(-\pi,\pi)$ for any $\w\o$ and $\nu$.
Hence they belong to  $\CC$. 
In addition, we have that  $\|G(\cdot,\w\o,\nu)\|_{\CC}\to +\infty$ as $\nu\to 0$. 

Let $\X_{\w\o} $ be the set of all processes  $x\in \ell_\infty$ with a single point 
spectrum degeneracy  at $\w\o$  compensating $G$, i.e.,   such that
\baaa
\|x\|_{\X_{\w\o} }\defi \sup_{\nu\in(0,1)}\|X G(\cdot,\w\o,\nu)\|_{\CC^*}<+\infty, \quad X=\F x.
\eaaa 
Let $\X_{\w\o} $ be the set of all processes  $x\in L_\infty(\R)$ such that
\baaa
\|x\|_{\X_{\w\o} }\defi \sup_{\nu\in(0,1)}\|X G(\cdot,\w\o,\nu)\|_{\CC^*}<+\infty, \quad X=\F x.
\eaaa 
We consider $\X_{\w\o} $ as a linear normed space with the corresponding norm.

In particular, this set included all processes with the any spectral gap $D\subset [-\pi,\pi]$ with non-empty interior  such that $e^{i\w\o}$ belongs to the interior of the arc $\{e^{i\o}\}_{\o\in D}$.

Let  $r\in (0,1)$ and be given. For all $\g>0$ , define 
\baa H_\g(z)\defi z\left(
1-\exp\left[-\frac{\g}{z+ 1-\g^{-r}}\right]\right).\label{wK}\eaa
We have that  $H_\g(e^{i\cdot})\in \WO(-\pi,\pi)\subset\CC$.

\begin{theorem}\label{ThP}  The functions 
  $\{H_{\g}\left( e^{i\cdot}\right)\}_{\g>0}\subset\CC$ are causal transfer functions defined   on $\ell_\infty$ such that, for any $\w\o\in (-\pi,\pi]$, \comm{REMOVE: for any set $x\in\X_{\w\o}$} there exists $\oo\g>0$ such that 
\baaa
\sup_{t\in\Z}|x(t+1)-\w x_\g(t)|\le\e\qquad \forall \g\ge \oo\g
\eaaa
 for  any $x\in\X_{\w\o}$ such that $\|x\|_{\X_{\w\o}}\le 1$.
Here
\baaa
\w x_\g (t)=e^{i(\w\o-\pi)  t}\sum_{s=-\infty}^t h_{\g}(t-s)e^{i(\pi-\w\o)  s}x(s), \quad h_\g=\F^{-1} H_{\g}.\qquad.
\eaaa
\end{theorem}
It will be shown below that the functions  $H_{\g}$  approximate  $e^{i\o}$ on $\TT$ for $\o\in (-\pi,\pi)$   as  $\g\to +\infty$, i.e., they represent a one-step predictor.

 These predictors were  introduced  in \cite{D12}. In \cite{D16}, some numerical experiments  for these predictors have been described, in particular, with different choices with choice of  $r$.

  \subsection{Recoverability in the case of unknown spectral gap}
For $\O>0$,   let $\U(\O)$ (or $\U_R(\O)$,  or $\U_I(\O))$ be the set of all signals $x\in\ell_\infty$
such that, for each $x$,  there exists a Borel measurable set $D=D(x(\cdot))\subset  [-\pi,\pi]$
such that $\mes D\ge \O$ and 
 $x\in\V(D)$ (or $\V_R(D)$, or $\V_I(D)$, respectively). Clearly,  $\U(\O)\subset \U_R(\O)\cap \U_I(\O)$.
 
Let  $\lfloor r \rfloor$ denotes  the integer part of $r>0$, and  $|\M|$ denotes the number of elements of  a set $\M$.

\begin{theorem}\label{ThRR} For any finite set  $\M\subset\ZZ$,  for any  
$x\in \U_R(\O)\cup \U_I(\O))$, for a given set  of observations 
$x(t)|_{t\in\ZZ\setminus\M}$, the number of possible different ordered sets  
$\{x(t)\}_{t\in\M}$ cannot exceed  $N\defi \lfloor 2\pi/\O\rfloor$.
\end{theorem}   

It can be noted that, in Theorem \ref{ThRR},  the estimate for the possible number of values in $\C^{|M|}$ for the non-observable vector $\{x(t)\}_{t\in\M}$ does not depend on $\M$ or $|\M|$.

In particular, if $\O>\pi/2$ then the vector $\{x(t)\}_{t\in\M}$ is uniquely defined by the observations
$x(t)|_{t\in\ZZ\setminus\M}$. If $\O>\pi/4$ then the vector $\{x(t)\}_{t\in\M}$ can take no more  than two possible values in $\C^M$ for any given  set of the observations
$x(t)|_{t\in\ZZ\setminus\M}$.

\section{Proofs}
 \par
   {\em Proof of Proposition \ref{prop1}}.
Let $\w f_k$, $\w g_k$, and $\w h_k$, be the Fourier coefficients for $f,g$,and $h$. 

We have that 
\baaa
\|h\|_{\CC}=\sum_{k\in\ZZ}|\w h_k|=\sum_{k\in\ZZ}\left|\sum_{d\in\ZZ}\w f_{k-d}
\w g_d\right|\le 
\sum_{k\in\ZZ}\sum_{d\in\ZZ} | \w f_{k-d}|\,|\w g_d|=\sum_{d\in\ZZ}
\,|\w g_d|\sum_{k\in\ZZ} | \w f_{k-d}|=\|f\|_{\CC}\|g\|_{\CC}.
\eaaa
This proves statement (i).

Further, let $f\in \WO(-\pi,\pi)$. 
We have that
\baaa
\sum_{k\in\ZZ}(1+k^2)|\w f_k|^2\le C { \|f\|_{\WO(-\pi,\pi)}^2}
\eaaa 
for some $C>0$ independent on $f$. Further, we have that
$df(\o)/d\o\in L_2(-\pi,\pi)$  and
\baaa
\sum_{k\in\ZZ}|\w f_k|\le {\left(\sum_{k\in\ZZ}(1+k^2)|\w f_k|^2 \cdot \sum_{k\in\ZZ}(1+k^2)^{-1}\right)^{1/2}
\le \|f\|_{\WO(-\pi,\pi)}\left( \sum_{k\in\ZZ}(1+k^2)^{-1}\right)^{1/2}.}
\eaaa
This proves statement (ii). $\Box$
\par
{\em Proof of Lemma \ref{lemma1}}. Let  $U_\CC\defi \{f\in\CC:\ \|f\|_{\CC}\le 1\}$.
For any $f\in U_\CC$, we have that 
\baaa
\left| \langle X_m,f\rangle\right|=\left| \sum_{t=-m}^m x(t)\w f_t\right|\le \max_{t:\ |t|\le m} |x(t)| \sum_{t=-m}^m |\w f_t|\le \|x\|_{\ell_\infty}.
\eaaa
Here $\w f_t$ are the Fourier coefficients for $f$.

For $r>0$, let 
$P(r)\defi\{ X\in\CC^*:\  |\langle X,f\rangle| \le r\quad \forall f\in U_\CC\}$.
We have that $\CC$ is a separable Banach space. By the Banach–Alaoglu theorem,  $P(r)$ is sequentially compact in the weak* topology of the dual space $\CC^*$ for any $r>0$; see, e.g., Theorem 3.17 \cite{Rudin}, p.68. Hence 
there exists a sequence  of positive  integers $m_1<m_2<m_3<...\,$ such that
 the subsequence  $\{X_{m_k}\}_{k=1}^\infty$ of the sequence $\{X_m\}_{m=1}^\infty \subset P(\|x\|_\infty)$ has a  weak* limit in $X\in P(\|x\|_\infty)$. 

Further, for any $f\in U_\CC$ and any integers $n>m>0$, we have that 
\baa
\left| \langle X_m-X_n,f\rangle\right|=\left| \sum_{t:\ m\le |t|\le n} x(t)\w f_t\right|\le \|x\|_{\ell_\infty} \!\!\! \sum_{t:\ m\le |t|\le n} |\w f_t|\to 0\quad \hbox{as}\quad m\to +\infty.
\label{Cauchy}\eaa

Let us prove that the original sequence $\{X_m\}$ also  has a weak* limit $X$ in $\CC^*$.

Let $f\in\CC$  be given. Let us show that for any  $\e>0$ there exists $N=N(f,\e)$ such that \baa
&&\left| \langle X_{m}-X,f\rangle\right|\le \e\quad \forall k\ge N.
\label{eps}\eaa
Since $X$ is a weak* limit of the 
subsequence  $\{X_{m_k}\}_{m=1}^\infty$, and  by the property 
(\ref{Cauchy}), it follows that for any  $\e>0$ there exists $N=N(f,\e)$ such that \baaa
&&\left| \langle X_{m_k}-X,f\rangle\right|\le \e/2 \quad\forall k\ge N,\\
&&\left| \langle X_{m}-X_{m_k},f\rangle\right|\le \e/2 \quad \forall k\ge N, \ \forall m>m_k.
\eaaa
Hence (\ref{eps}) holds. Hence the sequence $\{X_m\}$  has the same as  $\{X_{m_k}\}_{k=1}^\infty$ weak* limit $X$  in $\CC^*$  that belongs to $P(\|x\|_{\infty})$, i.e.,
 $|\langle X,f\rangle|\le \|x\|_\infty$ for all $f\in\CC$ such that $\|f\|_\CC\le 1$. 
 
 Furthermore, let a sequence  $\{t_k\}_{k=1}^\infty \subset\ZZ$ be such 
that $\lim_{k\to +\infty}|x(t_k)|=\|x\|_{\infty}$. Consider functions  $f_k(\o)=e^{i\o t_k}$; they belong to  $\CC$, and $\|f_k\|_\CC=1$ for all $k$.
We have that $|\langle X,f_k\rangle|=|x(t_k)|\to \|x\|_{\ell_\infty}$ as $k\to+\infty$. Hence 
$\sup_k|\langle X,f_k\rangle|\ge \|x\|_\infty$. It follows that the operator norm $\|X\|_{\CC^*}$ is $\|x\|_{\ell_\infty}$.
The proof that
$\langle X,f\rangle=\sum_{t\in\ZZ}x(k)\w f_k $ is straightforward.
This completes the proof. $\Box$
\par
{\em Proof of Theorem \ref{Th1}}.  We have that  the mappings $\GG:\CC^* \to \ell_\infty$ 
and $\F:\ell_\infty\to \CC^*$ 
are  linear and continuous.  


Let us show that  the mapping $\GG:\CC^*\to\ell_\infty$ is injective, i.e. that
\baaa
\hbox{if}\quad x=\GG(X)=0_{\ell_\infty}\quad \hbox{then}\quad X=0_{\CC^*}.
\label{injG}\eaaa  Suppose that  $x=\GG(X)=0_{\ell_\infty}$, i.e.  \baaa
x(t)=\langle X,e^{i\cdot t} \rangle=0
\eaaa
for all $t\in\ZZ$. In this case, for any $f\in\CC$, we have that $f=\sum_{k\in\ZZ}\w f_k e^{i\cdot k}$ for 
$\{\w f_k\}\in \ell_1$, and 
\baaa
\left|\langle X,f\rangle\right|=\left|\langle X,\sum_{k\in\ZZ}\w f_k e^{i\cdot k}\rangle\right|
\le \sup_{k\in\ZZ}\left|\langle X,e^{i\cdot k}\rangle\right| \sum_{k\in\ZZ}|\w f_k|=0.
\eaaa
This means that $X=0_{\CC^*}$.  Hence the mapping  $\GG:\CC^*\to\ell_\infty$ is injective.

 Let us show that  the mapping $\GG: \CC^*\to
\ell_\infty$ is surjective, i.e. that
$\GG(\CC^*)=\ell_\infty$.
Let $x\in\ell_\infty$ be any,  let $X_m(\o)=\sum_{t=-m}^m e^{-i\o t}x(t)$, and let $X=\F x$. We have that  $X\in\CC^*$. It can be calculated directly that $\langle X_m,e^{i\cdot t}\rangle =x(t)\Ind_{|t|\le m}$ for any $t\in\ZZ$.  Hence \baaa
x(t)
=\lim_{m\to +\infty}\langle X_m,e^{i\cdot t}\rangle=\langle X,e^{i\cdot t} \rangle 
\eaaa
for any $t\in\ZZ$.
Hence $x=\GG X$. Therefore, 
the mapping $\GG$ is surjective. Moreover, this proof implies  also that $\GG (\F x)=x$ for any $x\in\ell_\infty$.  In its turn, this implies that  $\F (\GG X)=X$ for all  $X\in\CC^*$.  

Hence   the mapping $\GG: \CC^*\to
\ell_\infty$  is a bijection,  $\GG^{-1}=\F$, and $\F^{-1}=\GG$. 
  
As was mentioned above, the continuity of the mapping $\GG: \CC^*\to
\ell_\infty$ is obvious. The continuity of the mapping  $\F=\GG^{-1}: \ell_\infty\to \CC^*$ 
follows from Lemma \ref{lemma1}; alternatively,  it can be shown  using e.g. Corollary 2.12(c) in \cite{Rudin}, p. 49.
This completes the proof  of Theorem \ref{Th1}. $\Box$

 \par
   {\em Proof of Lemma \ref{lemmaTF}}.  Let $x_m(t)=x(t)\Ind_{|t|\le m}$,  $X_m(\o)\defi \F x_m=\sum_{t=-m}^m e^{-i\o t} x(t)$.  We have that $X_m$ converses to $X$ in weak* topology of $\CC^*$ as $m\to +\infty$. Hence 
\baaa
&&\w x(t)= \langle \w X, e^{i\cdot t }\rangle=\langle X,H\left( e^{i\cdot }\right) e^{i\cdot t }\rangle=\lim_{m\to +\infty}\langle X_m,H\left( e^{i\cdot }\right) e^{i\cdot t }\rangle
\eaaa
for any $t\in\ZZ$. Here \baaa
\langle X_m,H\left( e^{i\cdot }\right) e^{i\cdot t }\rangle &=&\langle X_m,  \sum_{k\in\ZZ}\w h_k e^{-i\cdot k}e^{i\cdot t }\rangle
=\langle X_m, \sum_{k\in\ZZ}\w h_k e^{i\cdot (t-k)}
\rangle\\
&=&\langle \sum_{q=-m}^m e^{-i\o q} x(q) , \sum_{k\in\ZZ}\w h_k e^{i\cdot (t-k)}\rangle=\sum_{q=-m}^m x(q)\w h_{t-q},
\eaaa
since $q=t-k$ if and only if $k=t-q$. It follows that the sequence $\langle X_m,H\left( e^{i\cdot }\right)
e^{i\cdot t }\rangle$, $m=1,2,...$, has a limit  $\sum_{q\in \ZZ}x(q)\w h_{t-q}$ in $\C$.   This series  absolutely converges uniformly over any bounded set of  $x\in \ell_\infty$  since
\baaa
\left|\sum_{q\in \ZZ:\, |q|>m}x(q)\w h_{t-q}\right|\le \|x\|_{\ell_\infty}\sum_{q\in \ZZ:\, |q|>m} |\w h_{t-q}|\to 0\quad \hbox{as}\quad m\to +\infty.
\eaaa
This completes the proof.
$\Box$
\par
{\em Proof of Theorem \ref{ThC}}.   The assumptions on $H$ imply that
$H(z)=\sum_{k=0}^\infty \w h_k z^{-k}$,  i.e. that $H\left( e^{i\cdot }\right)=\sum_{k=0}^\infty\w h_k e^{-i\cdot k}$. Then the result follows from  Lemma \ref{lemmaTF}. $\Box$
 
In addition,  let us provide an alternative  proof of  Theorem \ref{ThC} that does not rely on Lemma \ref{lemmaTF}. {  Let $X\in \CC^*$ and $x=\F^{-1} X$ be such that $x(t)=0$ for $t<\tau$. For $m=1,2,...$, let 
 $x_m(t)=x(t)\Ind_{|t|\le m}$ and $X_m=\F^{-1} x_m=\sum_{t=-m}^m e^{-i\cdot t} x(t)$.
   Further, let $\w X_m\defi H(e^{i\cdot}) X_m$ and $\w x_m\defi \F^{-1}\w X_m$.
 From the standard theory of causal transfer functions for processes from $\ell_2$, we know that
 $\w x_m(\tau)=0$ for all $m$. As was shown in the proof of Theorem \ref{Th1}, we have that $X_m\to X$ as $m\to +\infty$ in weak* topology of $\CC^*$. Since $H(e^{i\cdot})f\in\CC$ for any $f\in\CC$, it follows that  $\w X_m =H(e^{i\cdot}) X_m \to \w X$ as $m\to +\infty$ in weak* topology of $\CC^*$. Hence $\w x_m(\tau)\to \w x(\tau)$ as $m\to +\infty$. Therefore, $\w x(\tau)=0$.} This completes the proof.
 $\Box$

\par
{\em Proof of Theorem \ref{ThR}}. 
Let  $x_1,x_2\in\V(D)$ be such that $x_1(t)=x_2(t)$ for $t\notin\M$.
Let $y\defi x_1-x_2$. It is easy to see that  $y\in\V(D)$.  
Furthermore, we have that  $y(t)=0$ for $t\notin\M$, hence  $Y(\o)=(\F y)(\o)=\sum_{t\in\M}e^{-it\o}y(t)$.  
Since a non-zero  finite combination of sine and cosine functions cannot  be identically zero on a interval, we can have that $y\in\V(D)$ only if $y(t)=0$ for any $t\in\M$, i.e. if $y=0$. This completes the proof.
$\Box$
\newpage

\par
{\em Proof of Theorem \ref{ThP}}. 
 The functions $H_\g(z)$  belong to $\CC$, and they
are analytic on $\DD=\{z: \ |z|>1\}$.
Hence they are causal transfer functions belonging to  $\CC$.


\par

Assume first that $\w\o=\pi$, i.e., $e^{i\w\o}=-1$.

Let \baaa
U_\g(z)\defi 1-\exp\left[-\frac{\g}{z+1-\g^{-r}}\right], \qquad V_\g(\o)\defi e^{i\o}U_\g\ew,
 \qquad z\in\C,\ \  \o\in(-\pi,\pi].
 \eaaa
 
In our setting, $x(t+1)$ is the output of anticausal convolution
with the transfer function $K(z)\equiv z$, i.e., $x(t+1)=\Z^{-1}(K\Z
x)(t)$.
 We have that 
\baaa
x(t+1)-\w x_\g(t)&=& \langle  X,[1-H_\g\eww]e^{i\cdot t}\rangle
= \langle   X, e^{i \cdot t}V_\g\rangle\\&
=&\langle   X, G(\cdot,\pi,\g^{-r})^{-1}G(\cdot,\pi,\g^{-r})^{-1}e^{i \cdot t}V_\g\rangle
\\&=&\langle   XG(\cdot,\pi,\g^{-r}),G(\cdot,\pi,\g^{-r})^{-1}e^{i \cdot t}V_\g\rangle.
\eaaa

\begin{lemma}\label{lemmaV}
\begin{itemize}
\item[(i)]  For any $\g>0$, the functions $zU_\g(z)$
are continuous on $\oo\DD$ and analytic on $\DD$. 
\item[(ii)]   $V(i\cdot)G(\cdot,\pi,\g^{-r})^{-1}\in\CC$ for any $\g>0$, and   
$\|V(i\cdot)G(\cdot,\pi,\g^{-r})^{-1}\|_{\CC}\to 0$ as $\g\to +\infty$.
\comm{$V_\g(i\o)\to 1$  as  $\g\to +\infty$ for all  $\o\in \R\setminus\{0\}$. 
$|V_\g\ew|G(\o,\pi,\g^{-r})^{-1}\le 1$ for any $\g\ge \g_0$ and $\o\in D(\g)$.}
\comm{\item[(iv)] There exists $\g_0>0$ and an open interval $D_0=(-\d,\d)\subset D$, $\d>0$, such that  
$|V_\g(i\o)|
\le 1$ for any $\o\in D_0$ and $\g\ge \g_0$.}
\ \end{itemize}
\end{lemma}
{\em Proof of Lemma \ref{lemmaV}.}
Clearly, $1-\exp(z)=-\sum_{k=1}^{+\infty}(-1)^k z^k/k!$ for $x\in\C$.  
Hence  \baaa
U_\g
(z) =-\sum_{k=1}^{+\infty}\frac{(-1)^k\g^k}{k!(z+1+\g^{-r})^k}
\eaaa
and    $V_\g \in  \WO (-\pi,\pi)\subset\CC$.

Since the growth of the module  $z$ on $\DD$  is being compensated by multiplying on $U_\g
(z)$, it  follows  that $zU_\g(z)$
are continuous and bounded on $\oo\DD$ and analytic on $\DD$. 
Then
statement  (i) Lemma \ref{lemmaV}  follows.

Further, we have that $\|V(i\cdot)G(\cdot,\pi,\g^{-r})^{-1}\|_{\CC}\le 
\|V(i\cdot)G(\cdot,\pi,\g^{-r})^{-1}\|_{\WO(-\pi,\pi)}$.
 To prove Lemma \ref{lemmaV} (ii), it suffices to show that 
$\|V(i\cdot)G(\cdot,\pi,\g^{-r})^{-1}\|_{\WO(-\pi,\pi)}\to 0$ as $\g\to +\infty$.

We have that $V_\g\ew G(\o,\pi,\g^{-r})^{-1}=e^{-\psi_\g(\o)}$, where
 \baaa
\psi_\g(\o)\defi \frac{c}{|e^{i\o}+1|^q+\g^{-r}}+\frac{\g}{e^{i\o}+1-\g^{-r}}.
\eaaa
The proof for Lemma \ref{lemmaV} (ii)  explores the fact that $\Re\psi_\g(\o)\to +\infty$ as $\g\to +\infty$ for all $\o\neq \pm\pi$ and  $\inf_\g\Re\psi_\g(\o)$  is bounded  in a neighbourhood of $\o=\pm\pi$. The proof 
 is  similar to the proof of Lemma 1 in \cite{D12} and is rather technical; it will be omitted here.
$\Box$

Furthermore, we have that
\baaa
|x(t+1)-\w x_\g(t)|&\le& \left\|XG(\cdot,\pi,\g^{-r})\right\|_{\CC^*} 
\left\|e^{i \cdot t}G(\cdot,\pi,\g^{-r})^{-1}V_\g\right\|_{\CC}\\ 
&=&\left\|XG(\cdot,\pi,\g^{-r})\right\|_{\CC^*} 
\left\|G(\cdot,\pi,\g^{-r})^{-1}V_\g\right\|_{\CC}\\
&\le&\sup_{\g>0}\|XG(\cdot,\pi,\g^{-r})\|_{\CC^*}\| G(\cdot,\pi,\g^{-r})^{-1}V_\g\|_{\CC}.\eaaa
By Lemma \ref{lemmaV} (ii),  the proof of Theorem  \ref{ThP} follows
for the case where $\w\o= \pi$. 
\par

For the case where   $\w\o\neq \pi$, we can apply these predictors $H_d$ to 
the signal $y(t)\defi e^{i(\pi-\w\o)  t} x(t)$. 
 Let $Y=\F^{-1} y$ and  $\w Y=H_d Y$;
this is a one-step prediction  process for $y(t)$, i.e. $\w y(t)\sim y(t+1)$. The implied one-step  prediction  process  $\w x$  for $x$ can be obtained as $\w x(t)=e^{i(\w\o-\pi)  t}\w y(t)$. This completes the proof of Theorem  \ref{ThP}. $\Box$

\par
{\em Proof of Theorem \ref{ThRR}}. Let $\w\V(D)$ and $\w\U(\O)$ denote either  
$\V_R(D)$ and $\U_R(\O)$ or $\V_I(D)$  and $\U_I(\O)$ respectively.
Suppose that   $x_1,x_2,x_2,...,x_{N+1}\in\w\U(\O)$ are such that $x_1(t)=x_2(t)=...=x_{N+1}(t)$ for all $t\notin\M$ and that at least  some of the vectors $v_k=\{x_k(t)\}_{t\in\M}$ are different.
By the assumptions, there exist Borel sets $D_1$,...,$D_{n+1}$  with the measure $\O$ or larger such that $x_k\in\w\V(D_k)$. Since  $N> \lfloor 2\pi/\O\rfloor$, we have that  $N+1> \lfloor 2\pi/\O\rfloor+1$. 
Hence there exist $m,n\in\{1,...,N+1\}$  such that
$\mes(D_n\cap D_m)>0$ and  $m\neq n$.

Let $y\defi x_m-x_n$. We have that  $y(t)=0$ for $t\notin\M$ and $y\in\V(D_n\cap D_m)$. The remaining part of the proof follows the proof of Theorem \ref{ThR}: we have 
$Y(\o)=(\F y)(\o)=\sum_{t\in\M}e^{-it\o}y(t)$. Since a non-zero  finite combination of the sine and cosine functions cannot be identically zero on a interval, we can have that $y\in\w\V(D_n\cap D_m)$ only if $y(t)=0$ for any $t\in\M$, i.e. if $y=0$. This contradicts supposition. This completes the proof.
$\Box$
 \section{Concluding remarks and further research}
 \begin{enumerate}
  \item   It would be interesting to characterise, in the  time domain, 
 the set of "irregular"  signals $x\in\ell_\infty$  such that  $X=\F x\in \CC^*\setminus C([-\pi,\pi])^*$.

    \item So far, it is unclear if the set of all predictable/recoverable processes 
 is everywhere dense in  the space  $\ell_\infty$ similarly to the space $\ell_2$, where the set of all band-limited processes is everywhere dense. 
 
  \item Theorem \ref{ThP}  implies that, for  any $x\in\ell_\infty$ such that $X=\F x$ 
 has a single point spectrum degeneracy of a certain kind,  and any $\tau\in \ZZ$,
 the observations of the one-sided tail $\{x(t)\}_{t\le\tau}$ defines the entire  signals $x$; therefore, this theorem implies as well the statement of Theorem \ref{ThR} for these processes. However, Theorem \ref{ThP}  does not cover  processes from wider classes  $\V_R(D)$ and $\V_I(D)$ covered by  Theorem \ref{ThR}.
\item  The proof of Theorem \ref{ThP}  implies that the functions $H_\g(e^{i\o})$ approximate $e^{i\o}$ for $\o\in (-\pi,\pi)$  as $\g\to +\infty$, in a certain sense.  
 \end{enumerate}

\end{document}